\documentclass[aps,prb,twocolumn,groupedaddress,showpacs,showkeys,floatfix]{revtex4}

\usepackage{graphicx}
\usepackage{amsmath}
\usepackage{bbm}
\usepackage{natbib}
\pdfoutput=1

\bibpunct[,]{(}{)}{,}{a}{}{,}

\begin{document}
\preprint{}

\title{Structure factors of harmonic and anharmonic Fibonacci chains\\
       by molecular dynamics simulations}

\author{Michael Engel}
\email{mengel@itap.uni-stuttgart.de}
\author{Steffen Sonntag}
\author{Hansj\"org Lipp}
\author{Hans-Rainer Trebin}
\affiliation{Institut f\"ur Theoretische und Angewandte Physik, Universit\"at
  Stuttgart, Pfaffenwaldring 57, 70550 Stuttgart, Germany}

\begin{abstract}
  
  The dynamics of quasicrystals is characterized by the existence of phason
  excitations in addition to the usual phonon modes. In order to investigate their
  interplay on an elementary level we resort to various one-dimensional model
  systems. The main observables are the static, the incoherent, and the
  coherent structure factor, which are extracted from molecular dynamics
  simulations. For the validation of the algorithms, results for the harmonic
  periodic chain are presented. We then study the Fibonacci chain with
  harmonic and anharmonic interaction potentials. In the {\em dynamic
    Fibonacci chain} neighboring atoms interact by double-well potentials
  allowing for phason flips. The difference between the structure factors of
  the dynamic and the harmonic Fibonacci chain lies in the temperature
  dependence of the phonon line width. If a bias is introduced in the well
  depth, dispersionless optic phonon bands split off.

\end{abstract}

\pacs{63.20.Ry, 61.44.Br, 02.70.Ns}
\keywords{Phonons; Phason flip; Quasicrystal.}

\maketitle

\section{Introduction}\label{sec:introduction}

\subsection{Motivation}

Quasicrystals are long-range ordered materials lacking translational
symmetry\cite{Intro}. Their diffraction patterns exhibit a dense set of sharp
Bragg reflections, that can be indexed by an integer linear combination of a
finite number of basis vectors which is larger than the dimension of space. As
a consequence the atomic configuration of quasicrystals is described with
reference to a higher-dimensional analog of a periodic lattice. Elementary
dynamic excitations within this `hyperspace' description are
phonons\cite{Quilichini97} and phasons \cite{Phasons}.

Phasons involve rearrangements of the structure by atomic jumps over short
distances, denoted `phason flips'. They are connected with many physical
properties of quasicrystals as for example elastic deformations
\cite{deBoissieu95}, dislocations \cite{Socolar86,
  Rosenfeld95}, diffusion \cite{Kalugin93, Bluher98}, and phase
transformations \cite{Steurer05}. Recently, indications for phason
flips\cite{Edagawa00, Abe03} have been observed by in situ transmission
electron microscopy. A coherent set of phason flips may form a 
static phason field, e.g.\ during a phase transformation or in the neighborhood
of a dislocation\cite{Engel05}.

By investigating the dynamics of quasicrystals one can find out the influence
of the quasiperiodicity on the phonon spectrum \cite{Quilichini97} and one may
gain a deeper understanding of phason flips\cite{Coddens00}. Both points can
be studied in x-ray or neutron diffraction experiments by measuring the
response of the system in frequency ($\omega$) and momentum ($q$) space.
Depending on the experimental setup, different functions can be obtained from
the scattering experiments: (1) The static structure factor $S(q)$ is the
usual -- not energy resolved -- diffraction image, measured with either x-rays
or neutrons. It is used for the determination of the atomic structure. (2) The
coherent structure factor\cite{DSF} $S(q,\omega)$ is studied via coherent
inelastic neutron scattering\cite{deBoissieu04} or alternatively via inelastic
x-ray scattering\cite{Krisch02}. It allows the determination of the phonon
dispersion relations. The experiments on icosahedral quasicrystals show well
defined acoustic phonon modes at small wave-vectors\cite{deBoissieu93} and
dispersionless broad optic bands at larger wave vectors\cite{Boudard95}. The
cross-over between the two regions is very sharp.  (3) The incoherent
structure factor\cite{DSF} $S_{\text{i}}(q,\omega)$ can be measured in
quasielastic neutron scattering.  Neutrons are exclusively used here, due to
the necessity of a high energy resolution. The technique also allows the
investigation of phason flips. In a series of experiments Coddens et al.
\cite{Lyonnard96, Coddens97, Coddens99, Coddens00} have found an anomalous
$q$-dependence of the quasielastic signal in icosahedral quasicrystals. They
interpreted it as correlated simultaneous jumps of several
atoms.\cite{Coddens99}

Up to now various calculations of the coherent structure factor of
quasiperiodic model systems have been published, see
Ref.~\onlinecite{Quilichini97}. Amongst them are the perfect one-dimensional
Fibonacci chain\cite{Ashraff89}, by static phason fields disordered Fibonacci
chains\cite{Naumis99}, and three-dimensional tilings\cite{Hafner93}. In these
studies the dynamical matrix is diagonalized, which is a purely analytic
method and yields the phonon dispersion relations only.  The results are
highly structured excitation spectra with a hierarchical system of
gaps\cite{Ashraff89}. The influence of anharmonicities, however, especially
the dynamics of phason flips has not been taken into account.

This `missing link' markes the starting point of our study. Here we present
calculations of the structure factors of special one-dimensional quasiperiodic
model systems by use of molecular dynamics (MD) simulations with either
harmonic potentials or potentials that allow for phason flips. Although
structure factors play such a central role in the dynamics of solids, not much
seems to be known about their exact forms for one-dimensional chains. Even for
the simple harmonic chain only few articles 
exist\cite{Yoshida81, Emery78, Florencio85, Radons85}.

\subsection{Model systems}

As a simple one-dimensional model for a quasiperiodic system we consider the
Fibonacci chain. It consists of particles arranged at two different distances:
large ones ($L$) and small ones ($S$). The length ratio $L/S$ equals the
number of the golden mean $\tau=\frac{1}{2}(\sqrt{5}+1)$. The sequence of the
distances is created recursively by the mapping $\lbrace L, S \rbrace \mapsto
\lbrace LS,L \rbrace$ with starting condition $L$. For example, after four
iterations the resulting sequence is $LSLLSLSL$.

\begin{figure}
  \centering
  \includegraphics[width=1.0\linewidth]{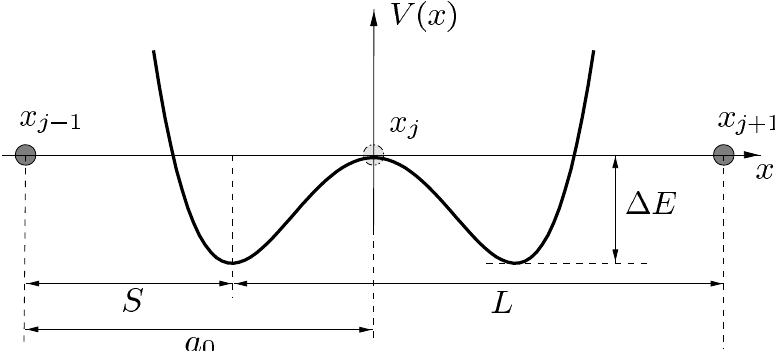}
  \caption{Double-well potential $V(x)=x^{4}-2x^{2}$ of the dynamic
    Fibonacci chain. The equilibrium distances are $S$ and
    $L$.\label{fig:doublewellpotential}}
\end{figure}

We want to study chains consisting of identical particles with
nearest-neighbor interactions. The Hamiltonian has the form
\begin{equation}
  \label{eq:hamiltonian}
  \mathcal{H} = \sum \limits_{j=1}^{N} \frac{p_j^2}{2}
    + V(x_{j} - x_{j+1} - a_{j}),
\end{equation}
where $x_{j}$ and $p_{j}$ are position and momentum of the $j$th particle.
The dynamic Fibonacci chain (DFC) is defined by the choices $a_{j}\equiv
a_{0}=\tau^{3}$ for the equilibrium distances and $V(x)=x^{4}-2x^{2}$ for the
interaction potential, respectively. The latter forms a double-well potential
with minima at $\pm1$ and a potential hill of height $\Delta E=1$ as shown in
Fig.~\ref{fig:doublewellpotential}. Because neighboring particles sit in
either of the potential minima, two nearest-neighbor distances $L=a_{0}+1$ and
$S=a_{0}-1$ are possible. They fulfil the constraint $L/S=\tau$ of the
Fibonacci chain.

The DFC shows two types of elementary excitations: Phonon vibrations in the
local minima and phason flips that interchange the particle distances $L$ and
$S$. At low temperatures only phonons are excited, phason flips have to be
activated thermally. With its neighbors at rest the activation energy of a
particle for a phason flip is $2\Delta E$.  This value is a result of the
perfect superposition of the potential hills of both neigbors.  The value is
lowered when the neighbors assist by stepping simultaneously to the inside or
outside during the phason flip thus creating a non-perfect superposition of
the potential hills. Since the particle distances $L$ and $S$ are
energetically degenerate, the total equilibrium potential energy is invariant
under a phason flip.

The occurrence of phason flips makes nonlinearity an intrinsic feature of
the DFC and an analytical treatment of the dynamics impossible. To understand
the influence of the nonlinearity, we study four model systems with increasing
complexity concerning their dynamical behavior:
\begin{itemize}
\item Harmonic periodic chain (HPC):\\ $V_{\text{HPC}}(x)=4x^{2}$ and
  $a_{j}=a=2\sqrt{5}$.
\item Harmonic Fibonacci chain (HFC):\\ $V_{\text{HFC}}(x)=4x^{2}$ and
  $a_{j}=L$ or $S$ according to the Fibonacci sequence.
\item Dynamic Fibonacci chain (DFC):\\ $V_{\text{DFC}}(x)=x^{4}-2x^{2}$ and
  $a_{j}=\tau^{3}$.
\item Asymmetric Fibonacci chain (AFC):\\
  $V_{\text{AFC}}(x)=V_{\text{DFC}}+\chi(x^{2}-1)^{2}(\epsilon x+x^{2}/2-1/2)$
  and $a_{j}=\tau^{3}$ with $\chi\in[0,1]$ and $\epsilon=\pm 1$.
\end{itemize}
The potentials of the HPC, HFC, and DFC are chosen to be identical in the
harmonic approximation around the equilibrium separation. The
average particle distance $a$ is the same for all four systems. In the case of
the Fibonacci chain the occurrence probabilities for L and S are given by
$\tau^{-1}$ and $\tau^{-2}$, hence $a=2\sqrt{5}$.

For the solution of the equations of motions we use a special molecular
dynamics (MD) code. The code is introduced in
Sec.~\ref{sec:dynamicstructurefactor} together with a short theoretical
background. The simplest system is, of course, the HPC. Exact solutions for
the equation of motion exist as a superposition of plane waves. We study the
dynamics of the HPC in Sec.~\ref{sec:HPC} as a reference system. The HFC
consists of particles arranged on the Fibonacci chain with distances $L$ and
$S$ interacting with the same harmonic potentials as the HPC, see
Sec.~\ref{sec:HFC}.  The DFC will then be studied in Sec.~\ref{sec:DFC}. In
the case of the AFC the particles in the two potential minima have different
eigenfrequencies. The parameters $\chi$ and $\epsilon$ determine the degree of
asymmetry. For more details we refer to Sec.~\ref{sec:AFC}. We finish with a
discussion and conclusion in Sec.~\ref{sec:conclusion}.

\section{Structure factors from molecular dynamics}\label{sec:dynamicstructurefactor}

\subsection{Definition of the structure factors}

We write the particle number density of the chain with $N$
particles as a sum of delta functions positioned along the particle
trajectories $x_{l}(t)$, $n({x},t) = \sum_{l=1}^{N} \delta(x - x_{l}(t))$. The
time dependent density-density correlation function and the density-density
autocorrelation function are defined as
\begin{subequations}
\begin{eqnarray}
  G(x,t) &=& \frac{1}{N}\int \left\langle
  n(x',t)n(x+x',0) \right\rangle \,\text{d}x' \nonumber\\
  &=& \frac{1}{N} \sum_{j,l}\left\langle \delta(x - x_{j}(t) +
         x_{l}(0)) \right\rangle,\\
  G_{\text{a}}(x,t) &=& \frac{1}{N} \sum_{l} \left\langle \delta(x - x_{l}(t) +
         x_{l}(0)) \right\rangle,
\end{eqnarray}
\end{subequations}
where the brackets denote the thermal average.\cite{CorrInterpretation} The
coherent and incoherent structure factor are the space-time Fourier
transforms\cite{Bee88},
\begin{subequations}
\label{eq:s_coh}
\begin{eqnarray}
\label{eq:s_coha}
  S(q,\omega) &=& \frac{1}{2\pi N}\int e^{-i\omega t}\sum_{j,l}{\left\langle
  e^{-iqx_{j}(t)}e^{iqx_{l}(0)}\right\rangle}\,\text{d}t,\nonumber\\\\
\label{eq:s_cohb}
  S_{\text{i}}(q,\omega) &=&  \frac{1}{2\pi N}\int e^{-i\omega
  t}\sum_{l}{\left\langle
  e^{-iqx_{l}(t)}e^{iqx_{l}(0)}\right\rangle}\,\text{d}t.\nonumber\\
\end{eqnarray}
\end{subequations}
Both functions are symmetric about $q=0$ and $\omega=0$. The static structure
factor is the integral of the coherent structure factor, $S(q)=\int
S(q,\omega)\,\text{d}\omega$, i.e.\ the Fourier transform of $G(x,0)$,
\begin{equation}
\label{eq:s_static}
S(q) = \frac{1}{N}\sum_{j,l}{\left\langle
    e^{-iqx_{j}(0)}e^{iqx_{l}(0)}\right\rangle}.
\end{equation}

\subsection{Molecular dynamics simulations}\label{ssec:MD}

For further calculations the particle trajectories are required as solutions
of the equations of motion. Since in the case of the anharmonic chains only
numerical solutions exist, we use a simple MD code. Initially the particles are placed on the
equilibrium positions of a finite chain of length $L$ with periodic
boundary conditions. The velocities are initialized according to a Gaussian
distribution. Its width determines the total energy and thus the temperature
of the system. The equations of motion are integrated by a
Verlet algorithm running for a simulation time $T_{\text{sim}}$. After
starting the simulation, the dynamics is not controlled by a thermostat or in
any other way.

For the direct numerical calculation of the Eqs.~(\ref{eq:s_coh}) we must
compute a fourfold sum: two sums over the particle number $N$ and two over the
time $T_{\text{sim}}$, one sum for the Fourier transform and one for the time
average. Note, that by assuming ergodicity the thermal average
$\left\langle\,\right\rangle$ can be replaced by a time average
$\frac{1}{T_{\text{sim}}}\int\,\text{d}t$ and additionally by an average over
several independent MD runs. For the sake of clarity the averaging over the MD
runs is suppressed in the following notation. We introduce a more compact
notation by defining the functions $f_{l}(q,t) = e^{i qx_{l}(t)}$. Let us
assume tentatively that these functions are periodic in time with period
$T_{\text{sim}}$ and in space with period $L$.  Then the Eqs.~(\ref{eq:s_coh})
and (\ref{eq:s_static}) are greatly simplified to
\begin{subequations}
\label{eq:SF_calc}
\begin{eqnarray}
\label{eq:SF_calca}
  S(q,\omega) &=& \frac{1}{2\pi NT_{\text{sim}}}\left\Vert\int e^{-i\omega t}
  \sum_{l} f_{l}(q,t)\,\text{d}t \right\Vert^{2}, \\
\label{eq:SF_calcb}
  S_{\text{i}}(q,\omega) &=& \frac{1}{2\pi NT_{\text{sim}}}\sum_{l}\left\Vert\int e^{-i\omega
  t} f_{l}(q,t)\,\text{d}t \right\Vert^{2},
\end{eqnarray}
\end{subequations}
and
\begin{equation}
S(q) = \frac{1}{NT_{\text{sim}}}\int\left\Vert
\sum_{l} f_{l}(q,t) \right\Vert^{2}\,\text{d}t.
\end{equation}
The equations differ in the order of the absolute square and the particle sum.
Since only two sums are left, an efficient numerical computation of the
structure factors is possible. Furthermore a fast Fourier transform is used
for the time integrals in Eqs.~(\ref{eq:SF_calc}).

It is left to discuss the periodicity conditions. The spatial periodicity
follows from the periodic boundaries used in the simulation.
Therefore the chain acts like a ring and the excitations can go round during the
simulation. To avoid such a behavior that would lead to unwanted
correlations, we limit the maximum simulation time by the quotient of the
length of the chain and the sound velocity $c_{s}$ to $T_{\text{max}} =
L/c_{\text{s}}$. For all the model systems HPC, HFC, and DFC the sound
velocity is the same, $c_{s}=a\sqrt{8}$, and $T_{\text{max}} = N/\sqrt{8}$. We
use $T_{\text{sim}}=T_{\text{max}}$.

The functions $f_{l}(q,t)$ are in general not periodic in time. There is no
reason, why the particles should be at the same positions at the end of the
simulation as at the beginning. To avoid this problem, we multiply
$f_{l}(q,t)$ with a window function $w(t)$ to enforce an artificial
periodicity.  The function $w(t)$ has to decrease fast enough -- both in
direct as in Fourier space -- towards the boundaries of its domains. We use a
normalized broad Gaussian function. Its width is chosen as large as possible
with the constraint that the Gaussian has decayed to a small enough value at
the interval boundaries. The effect of the Gaussian is a smoothing of the
structure factors by convolution with a narrow Gaussian. The exact value of
the width has no influence on the results.

\section{Harmonic periodic chain as reference system}\label{sec:HPC}

\subsection{Analytic calculations}

The harmonic periodic chain (HPC) is used as a reference system to test our
algorithms since its equations of motions can be solved analytically. If we
put $x_{l}(t)=u_{l}(t)+l a$, then the $u_{l}(t)$ are expressed by a
linear combination of normal modes. The wave vector $q$ and the frequency
$\omega$ are related according to the dispersion relation
$\omega(q)=2\omega_{0}|\sin(q a/2)|$.  Here, $w_{0}$ is the eigenfrequency of
a single particle. In the case of the model systems HPC, HFC, and DFC we have
$\omega_{0}=\sqrt{8}$.

For the HPC the thermal averages in the structure factors, Eqs.~(\ref{eq:s_coh})
can be calculated to be
\begin{subequations}
\label{eq:dynstruchpc}
\begin{eqnarray}
\label{eq:dynstruchpca}
S(q,\omega)&=&\frac{1}{2\pi}\int e^{-i\omega t} \sum_{l=-\infty}^{\infty}e^{-i
  q a l}\exp\left(-{\textstyle
  \frac{1}{2}}q^{2}\sigma_{l}^{2}(t)\right)\,\text{d}t, \nonumber\\
&&\\
\label{eq:dynstruchpcb}
S_{\text{i}}(q,\omega)&=&\frac{1}{2\pi}\int e^{-i\omega t}
\exp\left(-{\textstyle \frac{1}{2}}q^{2}\sigma_{0}^{2}(t)\right)\,\text{d}t.
\end{eqnarray}
\end{subequations}
where we used from the literature\cite{Yoshida81}
\begin{eqnarray}\label{eq:sigmahpc}
\sigma_{l}^{2}(t)&=& \left\langle[u_{l}(t)-u_{0}(0)]^{2}\right\rangle \nonumber\\
 &=& \frac{k_{B}T}{\omega_{0}^{2}}\left[l +
  \frac{1}{2}\int_{0}^{2\omega_{0}t}
  J_{2l}(s)(2\omega_{0}t-s)\,\text{d}s\right].
\end{eqnarray}
$J_{n}(s)$ is the Bessel function of the first kind of order $n$. The only
external parameter in these functions is the temperature $T$. The particle sum
and the Fourier transform have to be evaluated numerically. Due to the
translation invariance of the HPC the double sum of the Eqs.~(\ref{eq:s_coh})
is reduced to a single sum. The static structure factor of the HPC can also be
calculated analytically\cite{Emery78}
\begin{equation}\label{eq:dsfstathpc}
S(q)=\frac{\sinh(q^{2}\sigma^{2}/2)}
{\cosh(q^{2}\sigma^{2}/2)-\cos(qa)}, 
\end{equation}
where $\sigma^{2}=k_{B}T/\omega_{0}^{2}$. 

Let us take a closer look at the incoherent structure factor
$S_{\text{i}}(q,\omega)$. In the limit of small $T$ the term
$\exp(-\frac{1}{2}q^{2}\sigma_{0}^{2}(t))$ decays slowly with $t$ and we
substitute $\sigma_{0}(t)$ with its approximation for large $t$:
$\sigma_{0}(t)=|t| k_{B}T/\omega_{0}$ for $|t|\rightarrow\infty$. This leads
to a Lorentzian peak
\begin{equation}
S_{\text{i}}(q,\omega)=\frac{1}{\pi}\frac{\Gamma}{\Gamma^{2}+\omega^{2}},\quad
\Gamma=\frac{q^{2}k_{B}T}{2\omega_{0}}
\end{equation}
In the limit of large $T$ the term $\sigma_{0}(t)$ is approximated for small
$t$: $\sigma_{0}(t)=t^{2} k_{B}T$ for $|t|\ll \omega_{0}^{-1}$. Hence there is a
Gaussian peak
\begin{equation}
S_{\text{i}}(q,\omega)=\frac{1}{\gamma\sqrt{2\pi}}
\exp\left(-\frac{\omega^{2}}{2\gamma^{2}}\right) ,\quad \gamma=\sqrt{q^{2}k_{B}T}.
\end{equation}
The transition temperature between these two limiting cases is
$k_{B}T=4\omega_{0}^{2}/q^{2}$.

\begin{figure}
  \centering
  \includegraphics[width=1.0\linewidth]{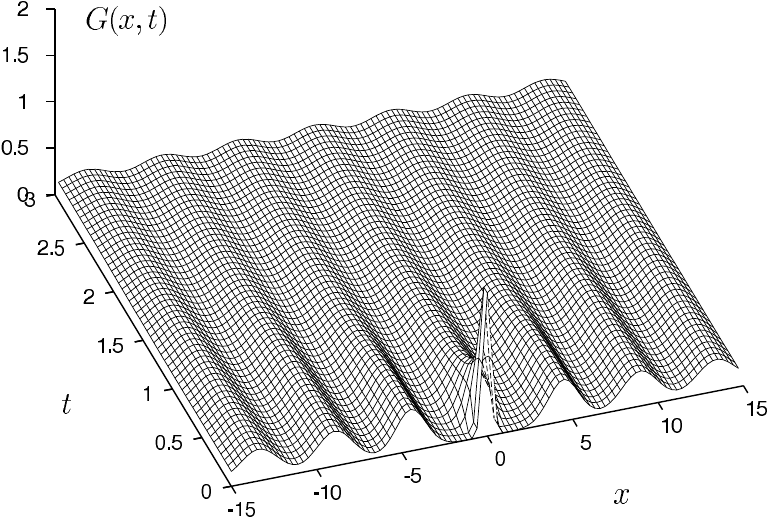}
  \caption{Density-density correlation function $G(x,t)$ of the HPC for
    $k_{B}T=0.5$. The Gaussians are centered at integer multiples of
    $x=a\approx 4.47$. \label{fig:correlationfunction}}
\end{figure}

The Fourier transform of the Eqs.~(\ref{eq:dynstruchpc}) yields the
correlation functions
\begin{subequations}
\label{eq:corrfunchpc}
\begin{eqnarray}
G(x,t)&=&\frac{1}{\sqrt{2\pi}}\sum_{l=-\infty}^{\infty}\frac{1}{\sigma_{l}(t)}
\exp\left(-\frac{(x+l a)^{2}}{2\sigma_{l}^{2}(t)}\right), \nonumber\\
\\
G_{\text{a}}(x,t)&=&\frac{1}{\sqrt{2\pi}}\frac{1}{\sigma_{0}(t)}
\exp\left(-\frac{x^{2}}{2\sigma_{0}^{2}(t)}\right).
\end{eqnarray}
\end{subequations}
The function $G(x,t)$ is shown in Fig.  \ref{fig:correlationfunction}. It
consists of a sum of Gaussians centered at the equilibrium positions of the
particles. The width of the Gaussians increases with temperature, as well as
with time and in space: $\lim_{x,t\rightarrow\infty}G(x,t)=\rho$ and
$\rho=1/a$. This means that there is no longe-range order. Indeed, for the
particle number density we have $\langle n(x)\rangle=\rho$, which is uniform
as in liquids \cite{Yoshida81}.  The autocorrelation function
$G_{\text{a}}(x,t)$ corresponds to the center peak at $x=0$.

\subsection{Simulation results}

\begin{figure}
  \centering
  \includegraphics[width=1.0\linewidth]{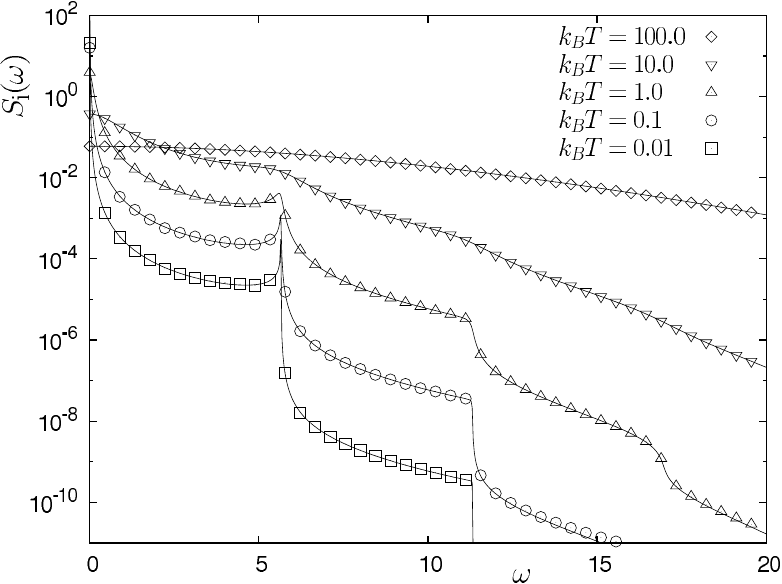}
  \caption{Incoherent structure factor $S_{\text{i}}(q,\omega)$ of the HPC for
  $q=\pi/a$, and different temperatures. The symbols mark the data from MD
  simulations with $N=1000$ particles, the lines the result from the
  analytical fomula, Eq.~(\ref{eq:dynstruchpcb}). The peaks and edges are at
  integer multiples of $\omega=2\omega_{0}=2\sqrt{8}$. \label{fig:sselfhpc}}
\end{figure}

In the case of the incoherent structure factor the variables $T$ and $q$ only
appear as combination $q^{2}k_{B}T$ in Eqs.~(\ref{eq:dynstruchpc}) and
(\ref{eq:sigmahpc}). Therefore it suffices to examine $S_{\text{i}}$ at a fixed wave
vector for different temperatures. We choose $q=\pi/a$ arbitrarily. The
results from MD simulation and the numerical integration of the analytical
formula, Eq.~(\ref{eq:dynstruchpcb}) are shown in Fig.~\ref{fig:sselfhpc} for
temperatures ranging from $0.01$ to $100.0$. There is a maximum at $\omega=0$,
called the quasielastic peak.  In the low temperature regime the maximum has a
Lorentzian shape.  Furthermore a one-phonon peak, a two-phonon edge, a
three-phonon edge, etc. are found at $2\omega_{0}\approx 5.7$,
$4\omega_{0}\approx 11.3$, $6\omega_{0}\approx 17.0$.\cite{xphonon} Note that
the multi-phonon contributions rapidly decay for larger $\omega$ (logarithmic
scale). At higher temperatures the curve smoothes and approaches a Gaussian
profile.

\begin{figure*}
  \centering
  \includegraphics[width=0.53\linewidth]{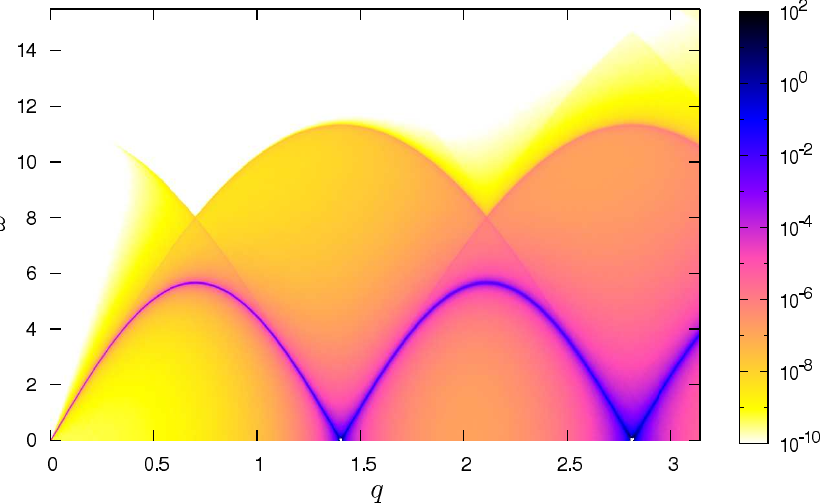}\hfill
  \includegraphics[width=0.45\linewidth]{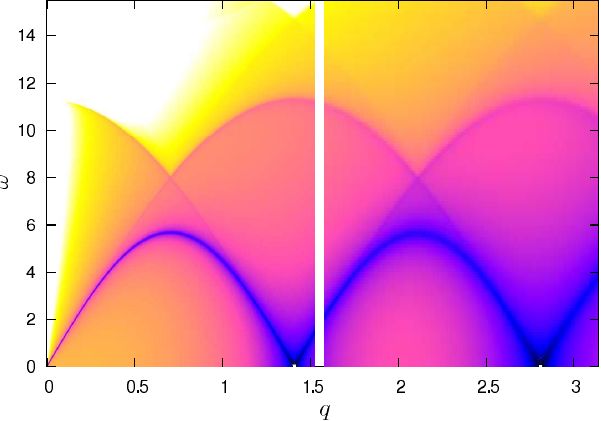}
  \caption{(Color online) Coherent structure factor $S(q,\omega)$ of the HPC
    with $N=6500$ particles from MD simulation. The temperatures are for
    $k_{B}T=0.01$ (a) and for $k_{B}T=0.1$ (b). One-, two- and three-phonon
    branches are observed.  They start at the reciprocal lattice points $2\pi
    n/a$. In (b) the output from the MD simulation (left side) is compared to
    the output of the analytical formula Eq.~(\ref{eq:dynstruchpca}) (right
    side). \label{fig:dsfcohhpc}}
\end{figure*}

The coherent structure factor $S(q,\omega)$ is shown in Fig.
\ref{fig:dsfcohhpc}. For a low temperature of $k_{B}T=0.01$ a one-phonon
branch, a two-phonon branch, and very weakly a three-phonon branch are
observed. The one-phonon branch has a Lorentzian line shape and follows the
phonon-dispersion relation. At higher $q$-values the branch broadens with a
width proportional to $k_{B}Tq^{2}$. This is similar to the temperature
behavior for the incoherent structure factor. The multi-phonon branches follow
the modified relations $\omega(q)=2n\omega_{0}|\sin(q a/2n)|$ with $n=2,3$.

In Fig.~\ref{fig:dsfcohhpc}(b) the comparison of the MD simulation (left side)
and the analytical formula Eq.~(\ref{eq:dynstruchpca}) (right side) is shown.
The temperature in this figure is $k_{B}T=0.1$, which is higher than in Fig.
\ref{fig:dsfcohhpc}(a). As a consequence the one-phonon branch is broader. For
both methods of calculating $S(q,\omega)$ a high accuracy over 12 orders of
magnitude is possible. The accuracy is only limited by the internal floating
point precision.

\begin{figure}
  \centering
  \includegraphics[width=1.0\linewidth]{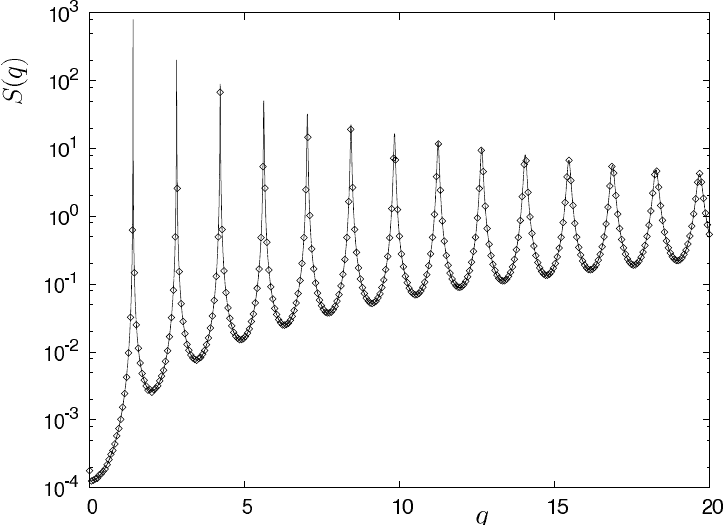}
  \caption{Static structure factor $S(q)$ of the HPC for $k_{B}T=0.02$. The
  symbols mark the data from a MD simulation with $N=1000$ particles, the line
    the result from the analytical fomula,
    Eq.~(\ref{eq:dsfstathpc}).\label{fig:dsfstathpc}}
\end{figure}

The static structure factor from Eq.~(\ref{eq:dsfstathpc}) is compared to the
results from the MD simulation in Fig.~\ref{fig:dsfstathpc}. $S(q)$ consists
of a sequence of Lorentzian peaks at the reciprocal lattice points. The
increasing width for larger wave vectors shows again that no long-range order
is present in the one-dimensional model system.

The MD simulations and analytical formulas show a perfect agreement. Therefore
we conclude that MD simulations are a well-suited numerical tool for
calculating the structure factors of the one-dimensional model systems.
Although we integrate the equations of motions with a good precision only on a
short time scale using the simple Verlet-algorithm, the statistics extracted
from the trajectories are correct.  This confirms our approach and encourages
us to proceed studying the phason dynamics of Fibonacci chains in the next
section.

\section{Harmonic Fibonacci chain: Influence of the quasiperiodicity}\label{sec:HFC}

By changing the interparticle equilibrium distances of the HPC to those of a
Fibonacci sequence with separations $L$ and $S$ we obtain the harmonic
Fibonacci chain (HFC). The interaction potential is left unchanged. Since the
incoherent structure factor is a function of the single particle motion only,
it does not depend on the equilibrium distances of the particles but only the
interaction potential.  Hence, the incoherent structure factor of the HPC and
of the HFC are identical. For the coherent structure factor the interparticle
distances become important. Instead of Eq.~(\ref{eq:dynstruchpca}) we now have
\begin{equation}
S(q,\omega)=\frac{1}{2\pi}\int e^{-i\omega t} \sum_{l}e^{-iqx_{l}^{0}}
\exp\left(-q^{2}\sigma_{l}^{2}(t)/2\right)\,\text{d}t.
\end{equation}
Here $x_{l}^{0}=\sum_{j=1}^{l}a_{j}$ for $l>0$,
$x_{l}^{0}=\sum_{j=l}^{-1}a_{j}$ for $l<0$ and $x_{0}^{0}=0$ denote the
equilibrium positions of the particles. For $k_{B}T=0$ this gives the Fourier
transform of the static Fibonacci chain
\begin{equation}
S(q,\omega)=\delta(\omega)\sum_{l}e^{-iqx_{l}^{0}}
\end{equation}
which is well known\cite{Levine86}. It consists of a dense set of
delta peaks with varying intensity, positioned at the reciprocal lattice
points
\begin{equation}
q=q_{0}\left(h+\tau h'\right),\qquad h,h'\in\mathbbm{N}
\end{equation}
with $q_{0}=\frac{2\pi}{a}\approx 1.40$.

\begin{figure*}
  \centering
  \includegraphics[width=1.0\linewidth]{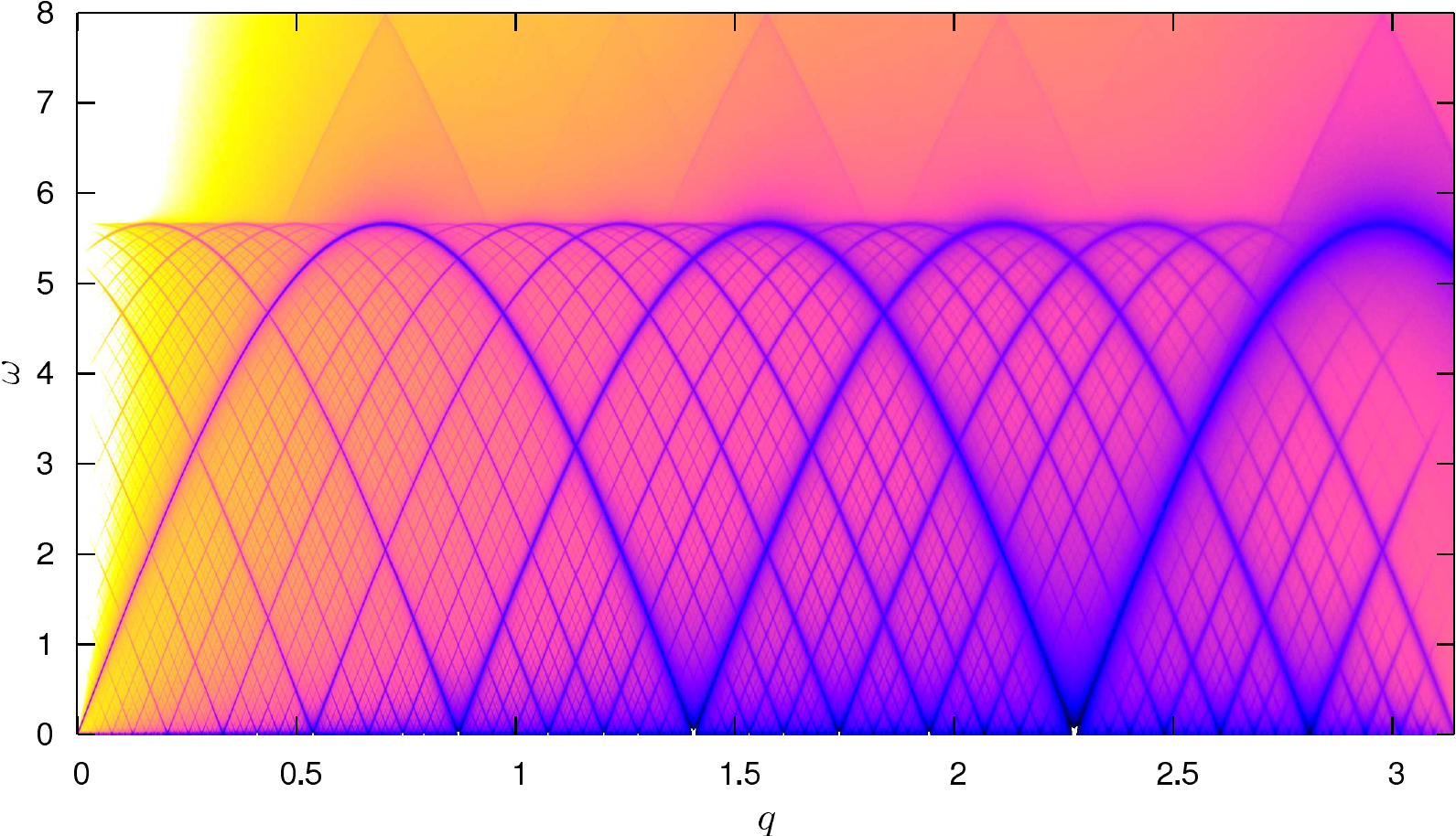}
  \caption{(Color online) Coherent structure factor $S(q,\omega)$ of the HFC
    with $N=13000$ particles for $k_{B}T=0.02$. It consists of a dense set of
    phonon branches starting from the reciprocal lattice
    points.\label{fig:dsfcohhfc}}
\end{figure*}

As shown in Fig.~\ref{fig:dsfcohhfc} for the temperature $k_{B}T=0.02$, the
coherent structure factor $S(q,\omega)$ of the HFC consists of many different
branches all following the one-phonon dispersion relation. Relative to each
other the branches are displaced in the $q$ direction. They start at the
reciprocal lattice points with the intensity of the respective delta peak.
Two-phonon branches are also found. The broadening of the branches
proportional to $k_{B}Tq^{2}$ has already been discussed for the HPC.

\begin{figure*}
  \centering
  \includegraphics[width=1.0\linewidth]{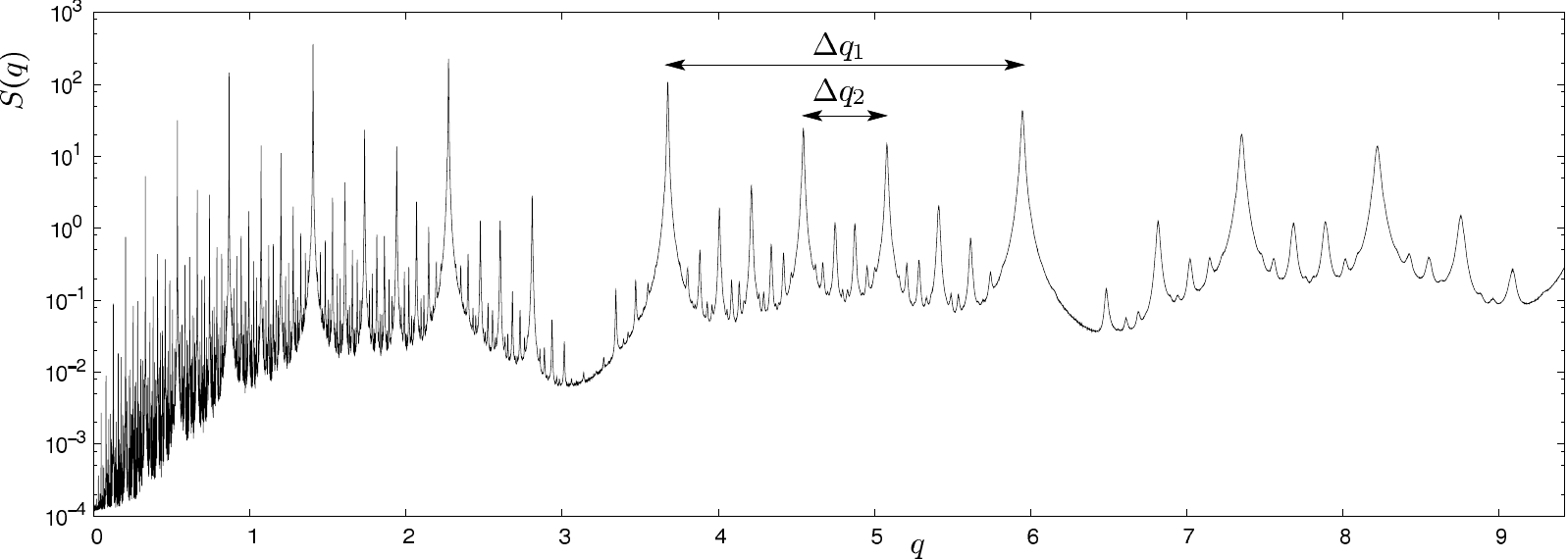}
  \caption{Static structure factor $S(q)$ of the HFC from a MD simulation with
  2000 particles at the temperature $k_{B}T=0.02$. \label{fig:sstathfc}}
\end{figure*}

From these findings and Fig.~\ref{fig:dsfstathpc} one can guess the form of
the static structure factor $S(q)$ for the HFC: Lorentzians are positioned at
the reciprocal lattice points with varying intensity. In
Fig.~\ref{fig:sstathfc} $S(q)$ is shown for $q\in[0,3\pi]$. For small $q$ a
large number of Lorentzians occur. With increasing $q$-value their width
increases and the stronger ones hide the weaker ones. Comparing the
position of the strong peaks and ignoring the change of their widths and
heights, there is a self-similarity in $S(q)$. The deflation factor is $\Delta
q_{1}/\Delta q_{2}=\tau^{3}$ as indicated in Fig.~\ref{fig:sstathfc}.

It is interesting to note that there are regions with very few peaks. They are
positioned around $q=\pi$, $q=2\pi$, etc. . The same regions are also special for
the coherent structure factor. As seen in Fig.~\ref{fig:dsfcohhfc} all the
weak one-phonon branches vanish towards $q=\pi$. Only the strong one-phonon
branch starting from the Bragg peak at $q=\tau q_{0}\approx 2.27$ remains.

\section{Dynamic Fibonacci chain: Occurrence of phason flips}\label{sec:DFC}

\subsection{Phason flips}

\begin{figure}
  \centering
  \includegraphics[width=1.0\linewidth]{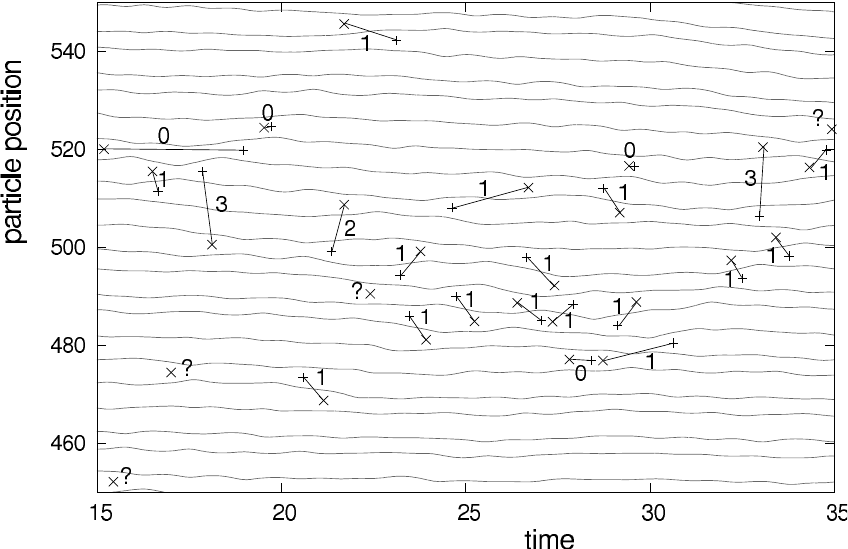}
  \caption{Snapshot of the particle trajectories of the DFC at the temperature
    $k_{B}T=0.6$. Changes in the particle distances from $L$ to $S$ and $S$ to
    $L$ are marked with a cross ($\times$) and a plus ($+$). A $L\rightarrow
    S$ change and a $S\rightarrow L$ change combined form a phason
    flip.\label{fig:direct}}
\end{figure}

Let us now proceed to the anharmonic chains with phason flips by investigating
the dynamic Fibonacci chain (DFC). It is built from identical particles that
interact with a symmetric double-well potential $V_{\text{DFC}}=x^{4}-2x^{2}$.
First the notion of a phason flip has to be specified. To do so we identify
the position of the changes from $L$ to $S$ and from $S$ to $L$ of the
interparticle distances along the particle trajectories. This is done in
Fig.~\ref{fig:direct}. Often a $L\rightarrow S$ change and a $S\rightarrow L$
change lie next to each other (particle distance 1) and the sequences $LS$ and
$SL$ are interchanged. But there are also many cases where the positions of
the two changes are separated by 0, 2, 3, or even more particle distances as
marked by lines in Fig.~\ref{fig:direct}. Sometimes it is not possible to find
a partner locally. Only in the long time average every $L\rightarrow S$ change
will eventually cancel with a $S\rightarrow L$ change.

In the literature on the Fibonacci chain a phason flip is understood as the
exchange of a $L$ and a neighboring $S$ particle distance. As we learned above
there are also other types of exchanges in the DFC. In the following we denote
by phason flip every pair of flip partners as those connected by lines in
Fig.~\ref{fig:direct}. Note that the times and positions of the phason flips
are not well defined.  Only their number can be estimated by
counting the changes in the particle distances as we will do now.

\begin{figure}
  \centering
  \includegraphics[width=1.0\linewidth]{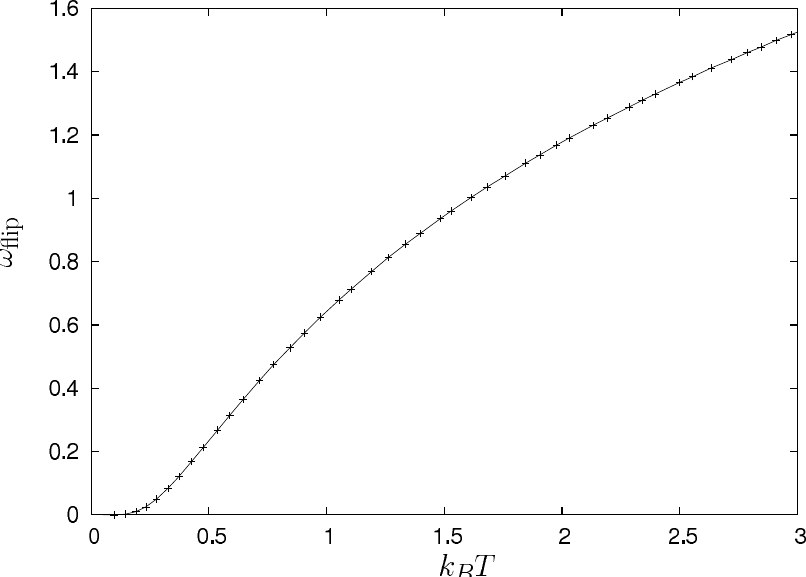}
  \caption{Average flip frequency as a function of the temperature
    $k_{B}T$. In the temperature range of the figure: $\omega_{\text{flip}}<
    \omega_{0}\approx 2.83$.
    \label{fig:av_flips}}
\end{figure}

The temperature dependence of the average phason flip frequency
$\omega_{\text{flip}}$ is shown in Fig.~\ref{fig:av_flips}. Phason flips start
to appear at about $k_{B}T=0.1$. At low $k_{B}T$ the average phason flip
frequency increases rapidly by thermal excitation and
$\omega_{\text{flip}}\ll\omega_{0}$. At higher temperatures $k_{B}T>0.4$ the
average phason flip frequency slowly saturates. In this region the
internal energy is comparable to the potential hill.

\subsection{Results for the structure factors}

\begin{figure}
  \centering
  \includegraphics[width=1.0\linewidth]{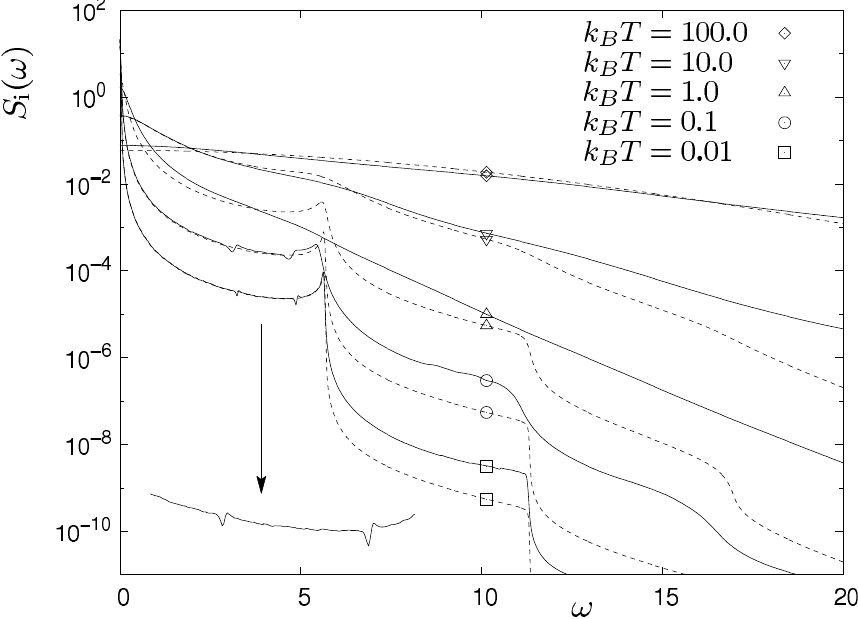}
  \caption{Incoherent structure factor $S_{\text{i}}(q,\omega)$ of the DFC
    (solid) and the HFC/HPC (dashed) for $q=\pi/a$ at different temperatures. The
    data was calculated using MD simulations with $N=1000$ particles.
    \label{fig:sselfdfc}}
\end{figure}

For the anharmonic chains no analytic results are available, in particular not
at elevated temperatures when phason flips occur. The incoherent structure
factors of the DFC and of the HFC/HPC differ remarkably and are shown in
Fig.~\ref{fig:sselfdfc} for $q=\pi/a$. The comparison leads to the following
conclusions:
\begin{enumerate}
\item At a fixed temperature, there are $\omega$-ranges where the curve for
  the DFC lies below the curve for the HFC/HPC and vice versa. Since we have
  $\int S_{\text{i}}(q,\omega)\,\text{d}\omega=1$ from Eq.~(\ref{eq:s_cohb}),
  the integral area between the two curves has to vanish.
\item At very low $k_{B}T$ and $\omega<2\omega_{0}\approx 5.7$ the curves of
  the DFC and the HFC/HPC cannot be distinguished in logarithmic scale except
  for two small bumps. They are a consequence of the anharmonicity of the
  interaction potential of the DFC and not related to the phason flips. At
  larger $\omega$ values the multi-phonon edges have different heights.
\item Above $k_{B}T=0.1$ the one-phonon peak and the multi-phonon edges in the
  curves for the DFC broaden and weaken considerably faster than in the curves
  for the HFC/HPC. They finally disappear at $k_{B}T=1.0$.
\item No additional peaks or edges occur at any temperature.
\end{enumerate}

Further MD runs show that different $q$-values change the temperature
dependence, but the general features remain unchanged: The phonon peaks and edges
broaden and weaken much faster with increasing temperature for the DFC than
for the HFC/HPC.

\begin{figure}
  \centering
  \includegraphics[width=1.0\linewidth]{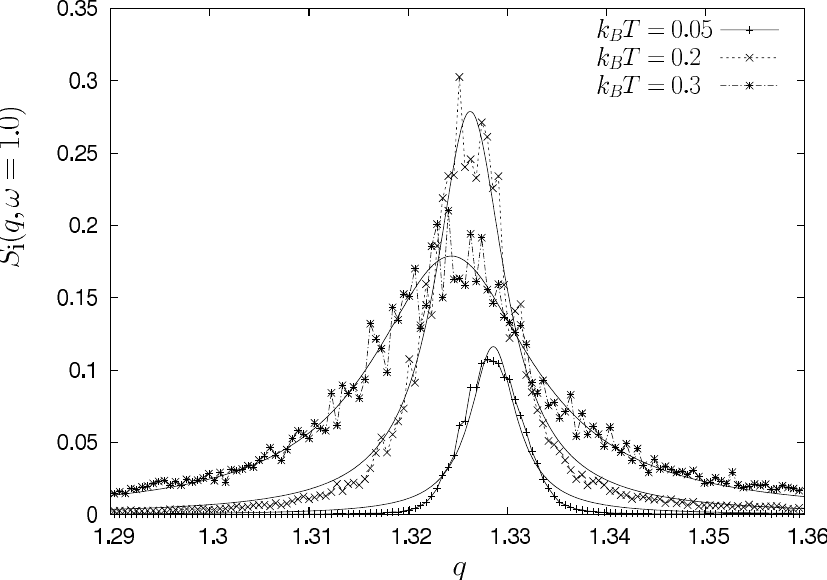}
  \caption{Cuts through $S(q,\omega)$ of the DFC for a fixed
    $\omega=1.0$ including a one-phonon peak. The solid curves are fits with
    Lorentzians. \label{fig:scohcut}}
\end{figure}

Similar conclusions follow for the \textit{coherent} structure factor of
the DFC. $S(q,\omega)$ for the DFC looks qualitatively similar to $S(q,\omega)$
for the HFC except that the branches broaden more quickly with
temperature. To compare the broadening let us look at plane cuts through
$S(q,\omega)$ for $\omega=1.0$ fixed. Three cuts for the temperature
$k_{B}T=0.05$, 0.2, and 0.3 are shown in Fig.~\ref{fig:scohcut}. A one-phonon
branch is located inside the cut resulting in a peak with approximate
Lorentzian line shape as indicated by the fits in the figure. Only for lower
temperatures the line shape deviates from a Lorentzian, which is seen at the
base of the peak for $k_{B}T=0.05$.

\begin{figure}
  \centering
  \includegraphics[width=1.0\linewidth]{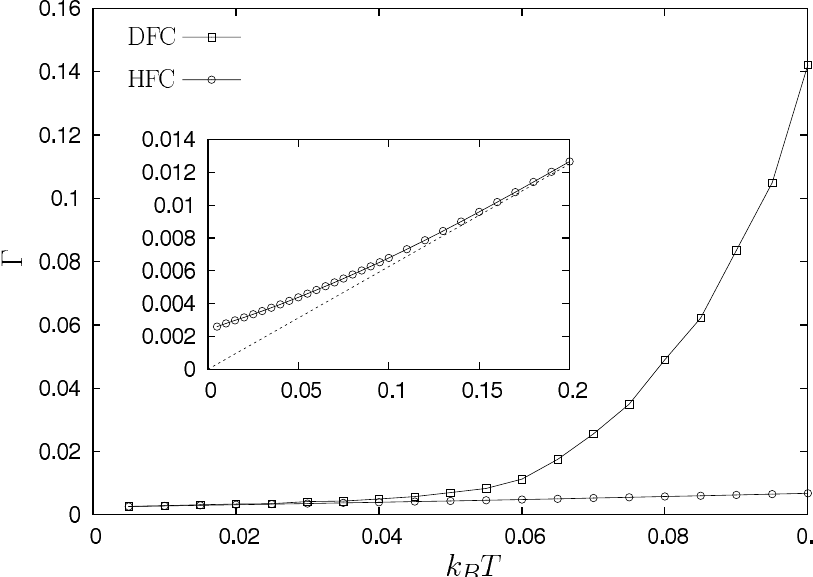}
  \caption{Width of the Lorentzian peak of Fig.~\ref{fig:scohcut} as a
    function of temperature for the HFC (inset) and the DFC.
    \label{fig:scoh_width}}
\end{figure}

The width of the Lorentzian in Fig.~\ref{fig:scohcut} is shown as a function
of temperature in Fig.~\ref{fig:scoh_width} for both, the HFC and the DFC. In
the case of the HFC the width increases linearly with temperature, as
discussed in Sec.~\ref{sec:HFC}. There is a deviation from linearity at
low temperatures. This is an artefact from the method of calculation from MD
simulation data. As explained in Sec.~\ref{ssec:MD} the structure factor is
convoluted with a Gaussian due to the finite simulation time. The (narrow)
Gaussian generates the observed offset. The convolution with the Gaussian is
also responsible for the shape of the cuts in Fig.~\ref{fig:scohcut} at low
temperature, deviating from the Lorentzian shape.

\begin{figure}
  \centering
  \includegraphics[width=1.0\linewidth]{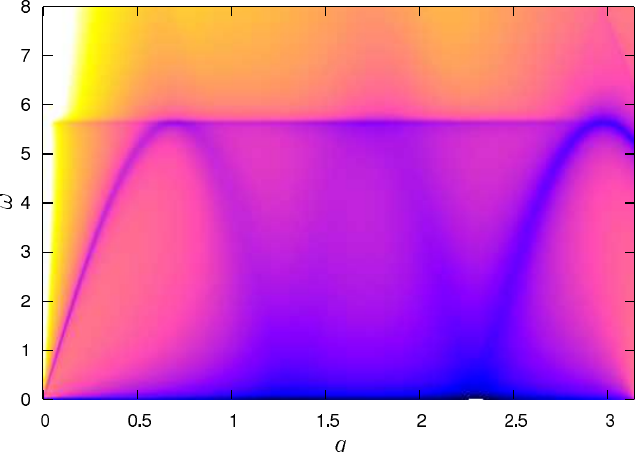}
  \caption{(Color online) Coherent structure factor $S(q,\omega)$ of a
    randomized Fibonacci chain with 2000 particles at the temperature
    $k_{B}T=0.02$. The same interaction potentials as in
    Fig.~\ref{fig:dsfcohhfc} have been used. \label{fig:scoh_random}}
\end{figure}

There is one aspect of the DFC that has been ignored up to now. Due to the
phason flips the original perfect quasiperiodic long-range order is slowly
decaying. With progressing simulation time the chain becomes randomized which,
however, has no effect on the incoherent structure factor. To
test the influence of the randomization on the \textit{coherent} structure
factor, a MD simulation was started with a 
totally randomized Fibonacci chain. The interaction potentials and the occurrence ratio of
$L$ and $S$ were not adapted. The result of the simulation is shown in
Fig.~\ref{fig:scoh_random}. Most of the details in $S(q,\omega)$ are lost by
the randomization process and the branches are strongly broadened, although
the simulation has been carried out at a low temperature of $k_{B}T=0.02$.

We summarize the results of this section: By the introduction of the
anharmonic double-well potential of the DFC the phonon peaks, edges, and
branches are strongly broadened and weakened with increasing temperature. There
are two effects responsible for the broadening: (1) The
anharmonicity and single phason flips affect the propagation of phonons. This
is seen in the incoherent structure factor. (2) The destruction of the
long-range order by a large number of phason flips. This is the main effect of
broadening for the coherent structure factor. The phason flips appear more or
less uniformly distributed over the simulation time and the particles.

\section{Asymmetric Fibonacci chain: Competing eigenfrequencies}\label{sec:AFC}

\subsection{Band gaps}

\begin{figure}
  \centering
  \includegraphics[width=1.0\linewidth]{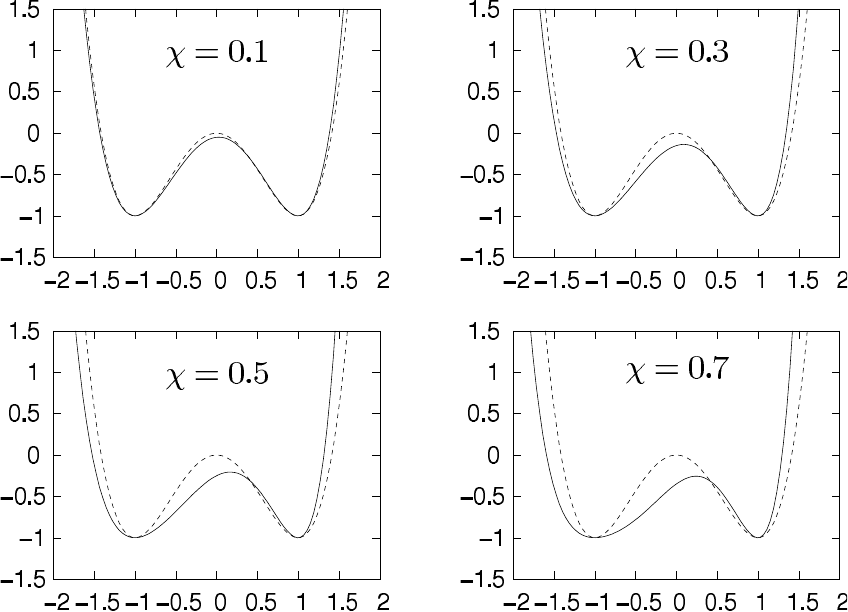}
  \caption{Interaction potential $V_{\text{AFC}}(x)$ of the AFC for different
    values of the parameter $\chi$ and $\epsilon=1$. $V_{\text{DFC}}$ is shown
    with dashed lines. \label{fig:doublewell2}}
\end{figure}

In the next step we modify the double well potential of the DFC. Like all the
previous model systems the asymmetric Fibonacci chain (AFC) is built of
identical particles, but they interact with the more complicated potential
$V_{\text{AFC}}=V_{\text{DFC}}+\Delta V$, see Fig.~\ref{fig:doublewell2}. The
additional term is given by
\begin{equation}
\Delta V(x)=\chi(x^{2}-1)^{2}(\epsilon x+x^{2}/2-1/2)
\end{equation}
with $\chi\in[0,1]$ and $\epsilon=\pm 1$. This term has been chosen in such a
way that the positions of the minima at $x=\pm 1$ are left invariant, but the
curvatures of the potential at those points is changed to
$V_{\text{AFC}}''(\pm1)=8(1\pm\epsilon \chi)$. In harmonic approximation a
particle will feel different eigenfrequencies, depending on the
nearest-neighbor configuration $SS$, $SL$, $LS$, or $LL$. In the case of
$\epsilon=1$ it is
$\omega_{\text{LL}}>\omega_{\text{LS}}=\omega_{\text{SL}}>\omega_{\text{\SS}}$
and for $\epsilon=-1$ it is
$\omega_{\text{LL}}<\omega_{\text{LS}}=\omega_{\text{SL}}<\omega_{\text{\SS}}$.
The sign change of $\epsilon$ corresponds to a mirror operation of
$V_{\text{AFC}}$ about the $y$-axis.

\begin{figure*}
  \centering
  \includegraphics[width=0.48\linewidth]{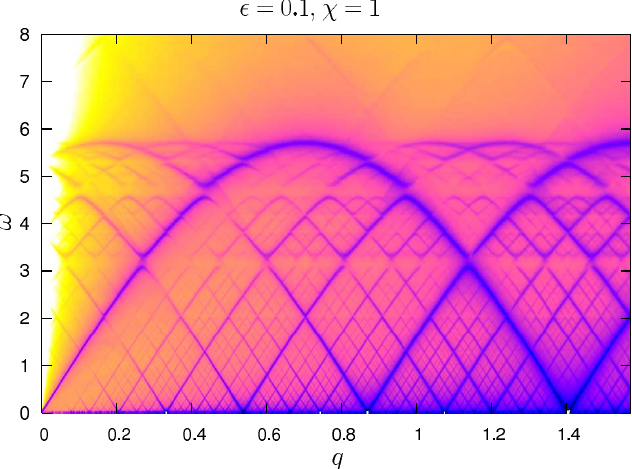}\qquad\vspace{0.5cm}
  \includegraphics[width=0.48\linewidth]{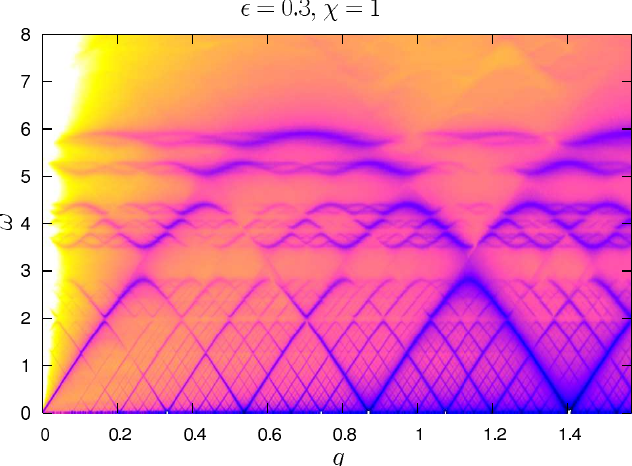}
  \includegraphics[width=0.48\linewidth]{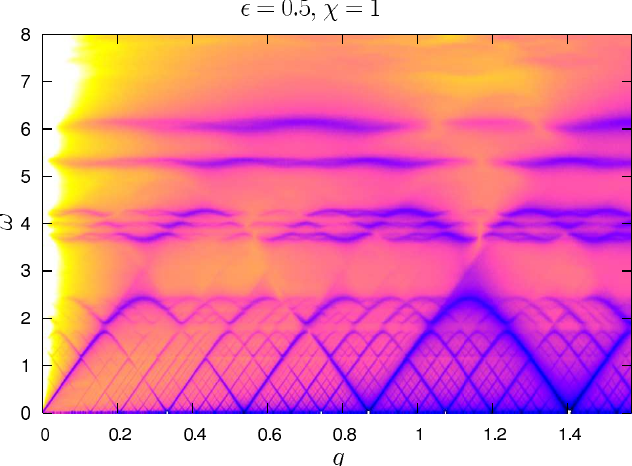}\qquad\vspace{0.5cm}
  \includegraphics[width=0.48\linewidth]{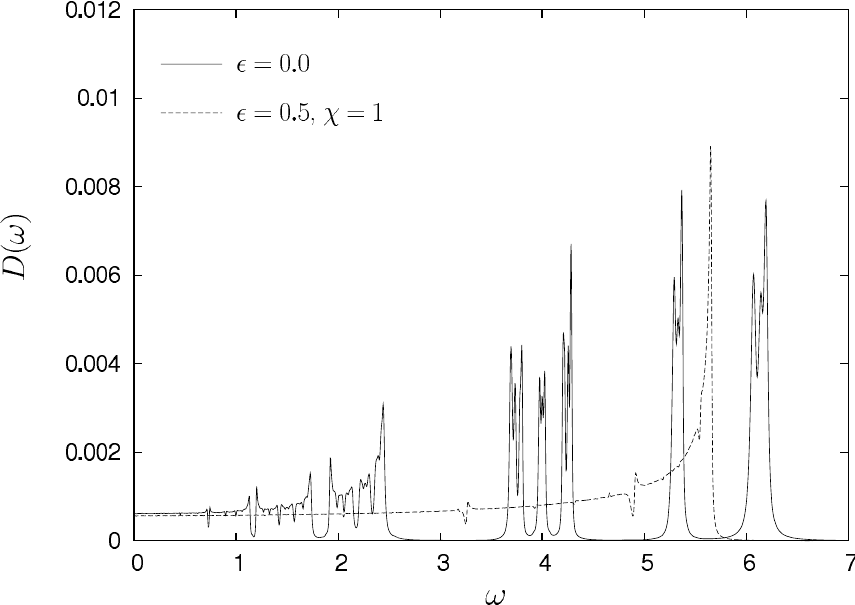}
  \includegraphics[width=0.48\linewidth]{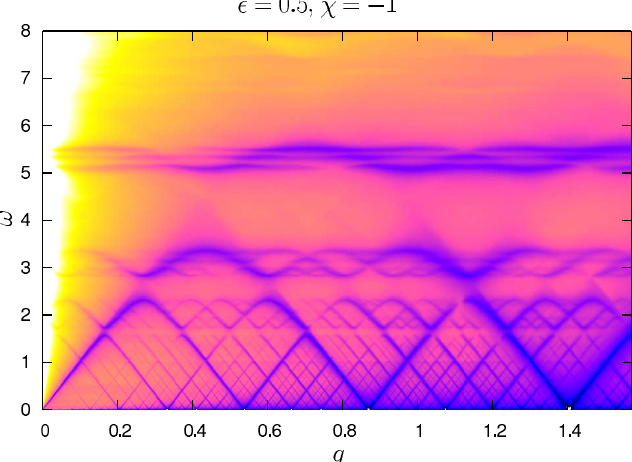}\qquad
  \includegraphics[width=0.48\linewidth]{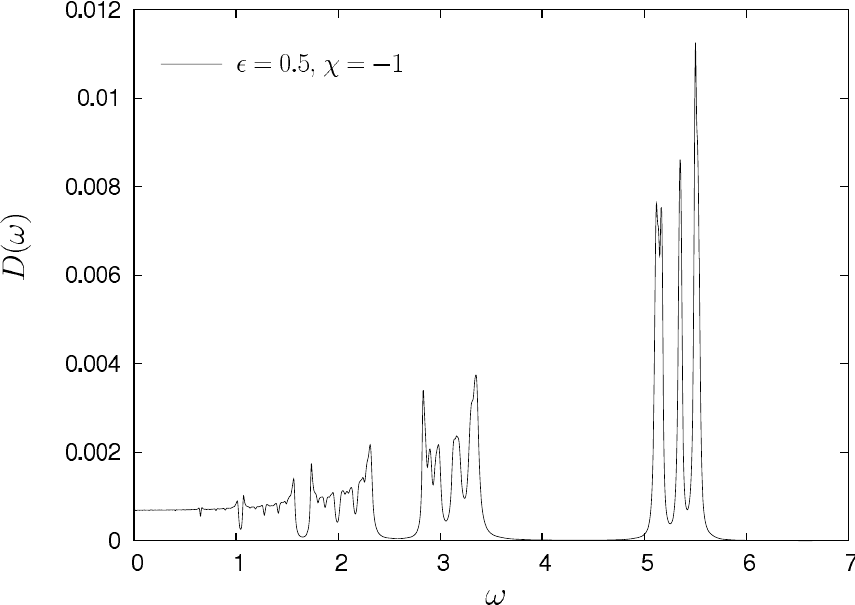}
  \caption{(Color online) Coherent structure factor $S(q,\omega)$ of the AFC
    for different values of $\chi$ and $\epsilon$ from MD simulations with
    6500 particles at $k_{B}T=0.01$. Band gaps are observed. The band gaps
    also appear in the DOS $D(\omega)$. \label{fig:scoh_afc}}
\end{figure*}

The coherent structure factor of the AFC with $\chi=0.1$, $0.3$, $0.5$ and
$\epsilon=\pm 1$ is shown in Fig.~\ref{fig:scoh_afc}. Band gaps of different
widths appear and broaden with increasing values for $\chi$. They are
positioned at the frequencies where one-phonon branches intersect each other.
Their positions and widths are different for $\epsilon=1$ and $\epsilon=-1$.
For $\epsilon=1$ three large gaps and several smaller gaps appear, whereas for
$\epsilon=-1$ only one very large gap, one medium gap and several small gaps
appear. The band gaps are a consequence of the competing eigenfrequencies due
to the asymmetric potential in the same way as band gaps appear in periodic
systems with several particles and eigenfrequencies per unit cell.

\subsection{Density of states}

The gaps are seen more clearly in the density of states (DOS) $D(\omega)$.
The DOS per particle is calculated from the velocity autocorrelation
function by a Fourier transform
\begin{equation}
D(\omega)=\frac{1}{\pi N}\int e^{-i\omega
t}\sum_{l}{\left\langle v_{l}(t)v_{l}(0)\right\rangle}\,\text{d}t,\quad
\omega\ge 0.
\end{equation} 
It is obtained from MD simulation data similar to the incoherent structure
factor: $f_{l}(q,t)$ has to be substited by $v_{l}(t)$ in
Eq.~(\ref{eq:SF_calcb}). By interchanging the Fourier transform and two time
derivatives we can alternatively write
\begin{equation}
D(\omega)= \frac{\omega^{2}}{\pi N}\int e^{-i\omega
t}\sum_{l}{\left\langle x_{l}(t)x_{l}(0)\right\rangle}\,\text{d}t.
\end{equation}
After a taylor expansion of the exponentials in Eq.~(\ref{eq:s_cohb}) a
connection to the incoherent structure factor is found,
\begin{equation}
D(\omega)=2\omega^{2}\lim_{q\rightarrow
  0}\frac{S_{\text{i}}(q,\omega)}{q^{2}},\quad\omega\neq 0.
\end{equation}

For comparison we present the DOS of a harmonic chain (HPC or HFC)
\begin{equation}
D(\omega)=\frac{2}{\pi}\frac{k_{B}T}{\sqrt{4\omega_{0}^{2}-\omega^{2}}}
\quad\text{for}\quad 0<\omega<2\omega_{0}
\end{equation}
and $0$ elsewhere. The total number of states is normalized to
$\int_{0}^{\infty}D(\omega)\,\text{d}\omega=k_{B}T$.

It can be checked, that the DOS of the harmonic chain fits well to the DOS of
the AFC for $\epsilon=0.0$, except small bumps that originate from the
anharmonicity of the potentials. In Fig.~\ref{fig:scoh_afc} the DOS of the AFC
for different values of $\chi$ and $\epsilon$ are drawn. The band gaps appear
at the same frequencies and the same widths as in the coherent structure
factor.

\section{Discussion and conclusion}\label{sec:conclusion}

At the end we would like to make some general remarks concerning phason flips
and phason modes: In the context of a hydrodynamic theory, phason flips can be
associated with phason modes. It was noted quite early\cite{Lubensky85}, that
phason modes are diffusive, in contrary to the propagating phonons. This means
that phason flips are only weakly coherent in space and in time, which of
course we have also observed here.  As a result their influence on the
structure factor is small making it difficult though still interesting to
study them by scattering experiments. Another point concerns the connection of
phason flips and quasiperiodicity. There is no reason why phason flips should
only occur in quasicrystals. Since interaction potentials are not sensitive on
the long-range order, phason flips in the form of atomic jumps can also occur
in periodic complex intermetallic phases, which is supported by recent
experimental results\cite{Dahlborg00, Dolinsek04}. In the case of our model
systems, it is equally possible to compare simulations of a periodic
$LSLSLS\ldots$ chain with harmonic and double-well potentials respectively.

In conclusion, we have investigated the dynamics of phonons and phason flips
in one-dimensional model systems with molecular dynamics simulations. An
efficient algorithm made it possible to calculate the structure factors with
high precision and in great detail. As a result multi-phonon contributions, --
although weak in comparison to the one-phonon peaks and branches -- have been
identified.  By introducing phasons in the model systems we were able to study
their influence on the structure factors, which is mainly a broadening of the
characteristic peaks, edges, and branches with temperature. The broadening can
be further split into a broadening due to the disorder as a result of
collective phason flips, i.e.\ a static phason field, and a broadening due to
the anharmonicity of the interaction potentials and single phason flips. The
work presented here is a first step.  Further studies on two-dimensional and
three-dimensional model systems with phason flips are under way.

\acknowledgments

We wish to thank T.\ Odagaki and M.\ Umezaki for stimulating discussions.


\begin{thebibliography}{20}
\bibitem{DSF}
The coherent structure factor $S(q,\omega)$ is also called `(total) scattering
function', `response function', or `dynamic(al) structure factor'. The
incoherent structure factor $S_{\text{i}}(q,\omega)$ is alternatively known as
`incoherent scattering function' and `self part of the structure factor'.
\bibitem{xphonon}
The notation `x-phonon' is adopted freely from the language of neutron
scattering. However in the case of a one-dimensional system there is no regular
phonon-expansion possible, as it is done for three-dimensional crystals.
\end{thebibliography}
\end{document}